\documentclass[prd,twocolumn,showpacs,floatfix,superscriptaddress,nofootinbib]
{revtex4}

\usepackage{graphicx}
\usepackage{epsfig}
\usepackage{bm}
\usepackage{amssymb}
\usepackage{float}
\usepackage{amsmath}
\usepackage{dcolumn}
\usepackage{cancel}
\usepackage[colorlinks]{hyperref}
\usepackage[usenames,dvipsnames]{color}
\hypersetup{
     breaklinks=true,
    pdfstartview={FitH},    
    colorlinks=true,       
    linkcolor=blue,          
    citecolor=red,        
    filecolor=magenta,      
    urlcolor=blue,           
    anchorcolor=green,      
    linktocpage=true
}

\providecommand{\U}[1]{\protect\rule{.1in}{.1in}}

\newcommand{\newc}{\newcommand}

\newc{\be}{\begin{equation}}
\newc{\ee}{\end{equation}}

\newc{\ba}{\begin{eqnarray}}
\newc{\ea}{\end{eqnarray}}
\newc{\bea}{\begin{eqnarray*}}
\newc{\eea}{\end{eqnarray*}}
\newc{\D}{\partial}
\newc{\ie}{{\it i.e.} }
\newc{\eg}{{\it e.g.} }
\newc{\etc}{{\it etc.} }
\newc{\etal}{{\it et al.}}
\newc{\lcdm}{$\Lambda$CDM }

\newc{\ra}{\Rightarrow}
 
\allowdisplaybreaks

\begin{document}

\title{Modified cosmology through Kaniadakis horizon entropy}

\author{Andreas Lymperis}\email{alymperis@upatras.gr} 
\affiliation{Department of Physics, University of Patras, 26500 Patras, Greece}

\author{Spyros Basilakos}
\email{svasil@academyofathens.gr}
\affiliation{Academy of Athens, Research Center for Astronomy $\&$ Applied Mathematics, Soranou Efessiou 4, 11-527, Athens, Greece }
\affiliation{National Observatory of Athens, Lofos Nymfon, 11852 Athens, Greece}

\author{Emmanuel N. Saridakis}
\email{msaridak@noa.gr}\affiliation{National Observatory of Athens, Lofos 
Nymfon, 11852 Athens, 
Greece}
\affiliation{CAS Key Laboratory for Researches in Galaxies and Cosmology, 
Department of Astronomy, University of Science and Technology of China, Hefei, 
Anhui 230026, P.R. China}
\affiliation{School of Astronomy, School of Physical Sciences, 
University of Science and Technology of China, Hefei 230026, P.R. China}


\begin{abstract}
We apply the gravity-thermodynamics conjecture, namely 
the first law of thermodynamics on the Universe horizon, but using the 
generalized 
Kaniadakis entropy instead of the  standard Bekenstein-Hawking one. The former 
is a one-parameter
generalization of the classical Boltzmann-Gibbs-Shannon entropy, arising 
from   a coherent and self-consistent relativistic statistical 
theory.
We obtain new modified 
cosmological scenarios, namely modified Friedmann equations, which contain new 
extra terms that constitute an effective dark energy sector depending on the 
  single model Kaniadakis parameter $K$. 
We investigate the 
cosmological evolution, by extracting analytical expressions for 
the dark energy density and equation-of-state parameters and we show that the 
Universe exhibits the usual thermal history, with a transition redshift from 
deceleration to acceleration at around 0.6. Furthermore, depending on the value 
of $K$, the dark energy equation-of-state parameter deviates from $\Lambda$CDM 
cosmology at small redshifts, while lying always in the phantom regime, and 
 at asymptotically large times the 
Universe always results in a dark-energy dominated, de 
Sitter phase. Finally, even in the case where we do not consider an 
explicit cosmological constant the resulting cosmology is very interesting and 
in agreement with the observed behavior.

\end{abstract}

\maketitle

\section{Introduction}

Observational data of the last two decades, reveal that the Universe has 
experienced   early-time and late-time acceleration stages. In this context and 
in order to explain this behavior, one can follow two main directions. The first 
direction is by constructing new modified and extended theories of gravity i.e. 
modify the left-hand-side of Einstein field equations by adding correction 
terms to the standard Einstein-Hilbert action (see for instance 
\cite{CANTATA:2021ktz,Capozziello:2011et,Cai:2015emx} and   references 
therein). This leads to modified classes of gravity such as $f(R)$ gravity 
\cite{Starobinsky:1980te,Capozziello:2002rd,DeFelice:2010aj,Nojiri:2010wj}, 
$f(G)$ gravity \cite{Nojiri:2005jg, DeFelice:2008wz}, Lovelock 
gravity \cite{Lovelock:1971yv, Deruelle:1989fj}, Weyl gravity
\cite{Mannheim:1988dj, Flanagan:2006ra} and Galileon theory 
\cite{Nicolis:2008in, Deffayet:2009wt, Leon:2012mt}. An alternate way is to 
start from the torsional formulation of gravity which leads to new modified 
extensions of gravity such as $f(T)$ gravity  
\cite{Ben09, Linder:2010py, Chen:2010va}, $f(T,T_G)$ gravity
\cite{Kofinas:2014owa,Kofinas:2014daa}, non-metricity 
\cite{Jimenez:2019ovq,Anagnostopoulos:2021ydo}, Finsler 
corrections 
\cite{Basilakos:2013hua} and other classes of geometrical modifications. The 
other direction is to modify the right-hand-side of Einstein field equations 
i.e. to introduce new matter fields such as the inflaton or the concept of 
dark energy 
\cite{Olive:1989nu,Bartolo:2004if,Copeland:2006wr,Cai:2009zp,Saridakis:2021qxb},
 providing new scenarios with extra degrees of freedom. 

Beyond the aforementioned directions in constructing new modified theories,  
there is a well-known conjecture that gravity can be expressed within laws of 
thermodynamics \cite{Jacobson:1995ab,Padmanabhan:2003gd,Padmanabhan:2009vy}. In 
particular, considering the Universe as a thermodynamical system, filled with 
   matter and dark-energy fluids and bounded by the apparent horizon 
\cite{Frolov:2002va,Cai:2005ra,Akbar:2006kj,Cai:2006rs}, the Friedmann 
equations 
can 
be expressed as the first law of thermodynamics. On the other hand, one can 
perform the reverse procedure, by applying the first law of thermodynamics on 
the Universe horizon and extract the Friedmann equations. The crucial point in 
applying the aforementioned conjecture in the context of modified theories, is 
that one should use the corresponding modified entropy relation which is valid 
in each modified theory 
\cite{Cai:2006rs,Akbar:2006er,Paranjape:2006ca,Sheykhi:2007zp,Jamil:2009eb,
Cai:2009ph,Wang:2009zv,Jamil:2010di, Sheykhi:2010wm,Sheykhi:2010zz,Gim:2014nba, 
Fan:2014ala,Lymperis:2018iuz,Sheykhi:2018dpn,Saridakis:2020lrg,Lymperis:2021hfk}
. Lastly, let 
us 
mention that new modified scenarios cannot be provided through the above 
procedure, due to the fact that since the modified entropy relation is 
needed, the modified theory needs to be known a priori.

On the other hand, several generalizations of the standard Boltzmann-Gibbs 
entropy and their cosmological implications have  been considered in the 
literature, such as Sharma-Mittal entropy \cite{sharma1975new}, R{\'e}nyi 
entropy \cite{renyi1961measures}, Shannon entropy \cite{Shannon:1948zz}, 
non-additive Tsallis entropy \cite{Tsallis:1987eu,Lyra:1998wz}, Barrow entropy 
\cite{Barrow:2020tzx}, etc, all of which   possess the standard 
 entropy as a particular limit. 
 
 One interesting such case of generalized entropy  is Kaniadakis 
  entropy \cite{Kaniadakis:2002zz,Kaniadakis:2005zk}. 
This is  a one-parameter
generalization of the classical Boltzmann-Gibbs-Shannon entropy, arising from  
 a coherent and self-consistent relativistic statistical 
theory, which  preserves the basic features of standard statistical theory, and 
recovers it in a particular limit. In such a framework the
corresponding distribution function is a one-parameter continuous deformation 
of the standard Maxwell-Boltzmann one. 

In the present work we are interested in  adopting the aforementioned 
reverse procedure,   using Kaniadakis entropy. In particular, we will apply 
the first law of thermodynamics in the Universe horizon, but using Kaniadakis 
entropy for the horizon entropy. In this way we obtain modified Friedmann 
equations, in which the 
new extra terms   will constitute 
the pillar for our investigation of the cosmological implications.

The plan of the manuscript is the following:  In Section \ref{model} we 
briefly review the application of the aforementioned conjecture in cosmology, 
and 
we present the new constructed modified scenario    arising from the
generalized Kaniadakis entropy instead of the usual Bekenstein-Hawking one. In 
Section \ref{cevol} we investigate the cosmological implications of the extra 
terms that appear in the modified Friedmann equations, focusing on the behavior 
of 
the dark energy density and equation-of-state parameters. Finally, in Section 
\ref{concl} we discuss our results.

\section{Modified cosmological scenario through Kaniadakis horizon entropy} 
\label{model}

We start our analysis by briefly reviewing the basic application of the first 
law of thermodynamics in the case of General Relativity,  and we extend our 
analysis by using the generalized Kaniadakis entropy instead of the standard 
one. Throughout the work we consider an expanding Universe filled with a matter 
perfect fluid, with energy density $\rho_m$ and pressure $p_m$, which is 
described by a 
homogeneous and isotropic Friedmann-Robertson-Walker (FRW) geometry with metric
\begin{equation}
ds^2=-dt^2+a^2(t)\left(\frac{dr^2}{1-kr^2}+r^2d\Omega^2 \right),
\label{metric}
\end{equation}
  where $a(t)$ is the scale factor, and with $k=0,+1,-1$ corresponding to flat, 
close and open spatial geometry respectively.

In order to apply the gravitational thermodynamics conjecture in cosmology,  
the 
first law 
is interpreted in terms of the heat, considered as the energy that flows 
through local Rindler horizons, applied on the horizon itself   
\cite{Jacobson:1995ab,Padmanabhan:2003gd,Padmanabhan:2009vy}, and in particular 
on the apparent horizon  
\cite{Bak:1999hd,Frolov:2002va,Cai:2005ra,Cai:2008gw}:
\begin{equation}
\label{apphor}
r_a=\frac{1}{\sqrt{H^2+\frac{k}{a^2}}},
\end{equation}
  where $H=\frac{\dot a}{a}$ the Hubble parameter and dots denoting derivatives 
with respect to $t$.
One then attributes to the Universe horizon an entropy and a temperature 
that arise from the corresponding relations of black hole thermodynamics. In 
  the case of General 
Relativity one applies the usual Bekenstein-Hawking entropy on the horizon, 
namely
\begin{equation}
\label{Horentropy}
S_{BH}=\frac{1}{4G} A,
\end{equation} 
where $A=4\pi r^{2}_{a}$ is   the area and
$G$ is the gravitational constant (we use the natural units $\hbar = k_{_B} = 
c =1$)  \cite{Padmanabhan:2009vy}.
On the other hand, 
 for the  horizon temperature we apply the standard relation   
 which   does not depend on 
the underlying 
gravitational theory \cite{Gibbons:1977mu}:
\begin{equation}
\label{Th}
 T=\frac{1}{2\pi r_a}.
\end{equation} 

 For a dynamical Universe, the heat flow through the horizon during a time 
interval $dt$ can be calculated to be \cite{Cai:2005ra}
$
\delta Q=-dE=A(\rho_m+p_m)H r_{a}dt.
$
Thus, the first law of thermodynamics reads
$
-dE=TdS$.
 Differentiation of (\ref{Horentropy}) immediately gives $dS=2\pi r_a \dot{r}_a 
dt/G$, where $\dot{r}_a$ can be obtained from (\ref{apphor}). Substituting 
everything in the first law we obtain
\be
\label{cFE1}
-4\pi G (\rho_m +p_m)= \dot{H} - \frac{k}{a^2}.
\ee 
  Furthermore, imposing the conservation equation for the matter fluid 
$
 \dot{\rho}_m +3H(\rho_m +p_m)=0,
$
 into (\ref{cFE1}) and integrating we obtain
\be 
\label{cFE2}
\frac{8\pi G}{3}\rho_m =H^2+\frac{k}{a^2}-\frac{\Lambda}{3},
\ee
  where $\Lambda$ is the cosmological constant, obtained as the integration 
constant. Hence, by applying the gravity-thermodynamics 
conjecture, we were able to obtain the Friedmann equations starting from 
the first law of thermodynamics. We mention here that we 
  imposed  the assumption that after equilibrium 
establishes, the Universe fluid acquires the same temperature with the horizon 
one, which is true for the late-time Universe
\cite{Padmanabhan:2009vy,Frolov:2002va,Cai:2005ra,Akbar:2006kj,Izquierdo:2005ku,
Jamil:2010di}.

As we mentioned in the Introduction,  the above procedure can be extended to 
modified gravity theories too, if one uses  the 
corresponding modified entropy of each theory 
\cite{Cai:2006rs,Akbar:2006er,Paranjape:2006ca,Sheykhi:2007zp,Jamil:2009eb,
Cai:2009ph,Wang:2009zv,
Jamil:2010di, Gim:2014nba, Fan:2014ala} instead of the general-relativistic
entropy relation (\ref{Horentropy}). Hence, one deduces that if we use 
the Kaniadakis entropy we will obtain novel modifications in the Friedmann 
equations. This will be done in the following, after a brief introduction to 
this extended entropy.

\subsection{Kaniadakis entropy} \label{kanentropy}

Kaniadakis entropy or K-entropy is a one-parameter
generalization of the classical Boltzmann-Gibbs-Shannon entropy, which arises 
from   a coherent and self-consistent relativistic statistical 
theory, which  preserves the basic
features of standard statistical theory, and recovers it in a particular limit 
\cite{Kaniadakis:2002zz,Kaniadakis:2005zk}. 
In the case of Kaniadakis generalized statistical theory the
corresponding distribution function is a one-parameter continuous deformation of 
the standard Maxwell-Boltzmann one. In particular, Kaniadakis entropy is given 
by 
\begin{eqnarray}
S_{K}=- k_{_B} \sum_i n_i\, \ln_{_{\{{\scriptstyle
K}\}}}\!n_i  ,
\end{eqnarray}
with $k_{_B}$ the Boltzmann constant,  
where $\ln_{_{\{{\scriptstyle
K}\}}}\!x=(x^{K}-x^{-K})/2K$, and
$-1<K<1$ is the dimensionless Kaniadakis parameter that quantifies the 
deviation from standard statistical mechanics, with the latter being recovered 
in the limit $K\rightarrow0$.
Within this
generalized theory the distribution function becomes 
\cite{Kaniadakis:2002zz,Kaniadakis:2005zk}
\begin{eqnarray}
 n_i= \alpha \exp_{_{\{{\scriptstyle K}\}}}[-\beta
(E_i-\mu)]  , 
\end{eqnarray}
with $\exp_{_{\{{\scriptstyle K}\}}}(x)=
\left(\sqrt{1+K^2x^2}+K x\right)^{1/K}$, 
 $\alpha=[(1-K)/(1+K)]^{1/2K}$, 
$1/\beta=\sqrt{1-K^2}\,\,k_{_{B}}\!T$, and where the chemical
potential $\mu$ can be fixed through normalization.  
Equivalently, Kaniadakis entropy    can be expressed as 
\cite{Abreu:2016avj,Abreu:2017fhw,Abreu:2017hiy,Abreu:2018mti,Yang:2020ria,
Sharma:2021zjx,
Abreu:2021avp,Drepanou:2021jiv}
\be \label{kstat}
S_{K} =-k_{_B}\sum^{W}_{i=1}\frac{P^{1+K}_{i}-P^{1-K}_{i}}{2K},
\ee
with  $P_i$ the probability of a system to be in  a specific microstate and $W$ 
 the total configuration number.

Applying the above in the case of  black holes (which will be the basis for the 
cosmological application), considering  that  $P_i=1/W$,  and  using the fact 
that Boltzmann-Gibbs entropy is $S\propto\ln(W)$, while   the 
Bekenstein-Hawking entropy is given by (\ref{Horentropy}), we obtain
  $W=\exp(A/4G)$, where from now on we use the natural units in which the 
Boltzmann 
constant   $k_{_B}$ is 1 \cite{Moradpour:2020dfm}. Hence, for the black hole 
application of   Kaniadakis entropy we obtain \cite{Moradpour:2020dfm}
 \be \label{kentropy}
S_{K} = \frac{1}{K}\sinh{(K S_{BH})},
\ee
 which for  $K\rightarrow 0$ recovers the standard Bekenstein-Hawking entropy, 
namely $S_{K\rightarrow 0}=S_{BH}$. We mention here that since the above 
expression is an even function, $S_{K}=S_{-K}$ and thus in the following we 
focus on the      $K\geq0$ region.

For completeness we give the relation of Kaniadakis entropy with other 
generalized entropies, such as the Tsallis one. In particular,  
the non-extensive Tsallis entropy $S^{T}_q$, where $q$ is the parameter that 
quantifies the deviation from Bekenstein-Hawking entropy 
\cite{Tsallis:1987eu,Tsallis:2012js}, is related to Kaniadakis entropy through
 \cite{Abreu:2017hiy,Moradpour:2020dfm,Nunes:2015xsa}
\be 
\label{Tsallisen}
S_{K} =\frac{S^{T}_{1+K}+S^{T}_{1-K}}{2}.
\ee
We mention here that there are two Tsallis entropies (equations (6) and 
(20) of \cite{Tsallis:2012js}). The first one is the 
Tsallis entropy used in (\ref{Tsallisen}), while the second one   leads to 
$S^{T}=\gamma A^\delta$, with $A$ the area and $\gamma$ and $\delta$ the two 
parameters. This second entropy does not satisfy  (\ref{Tsallisen}).
Nevertheless, there is another similar entropy, namely  Barrow entropy 
 $S^{B}_\Delta$,  which arises from 
  quantum-gravitational 
effects that impose intricate, fractal structure on the surface of the
black hole, where $\Delta$ is the parameter that 
quantifies the deviation from Bekenstein-Hawking entropy \cite{Barrow:2020tzx}. 
Barrow entropy is similar to the second Tsallis entropy and does not satisfy 
(\ref{Tsallisen}). In general, the free parameters in 
generalized entropies should be estimated by observations and experiments. Such 
entropies  are proper entropy measures for   complex systems, 
long-range interacting systems, and fractal systems.   Barrow's pioneering 
work shows that Tsallis non-extensive second entropy may also be explained in 
the quantum-gravitational framework, and thus that  gravity and its quantum 
features can provide a more enlightening picture of the non-extensivity 
\cite{Mejrhit:2020dpo,Mejrhit:2019oyi,Ourabah:2019mly,Moradpour:2019yiq,
Shababi:2020evc}. Hence, based on the bounds of the $\Delta$ parameter 
of Barrow entropy \cite{Barrow:2020kug} we may acquire a better understanding of 
Tsallis second entropy and its free parameters.

\subsection{Modified Friedmann equations through Kaniadakis entropy} 
\label{mfekentropy}

We can now proceed in applying the gravity-thermodynamics approach described 
above,  but instead of the standard Bekenstein-Hawking entropy relation we will 
use the generalized Kaniadakis entropy, namely equation (\ref{kentropy}). In 
particular, differentiating  (\ref{kentropy}) we    acquire
\be \label{dsk}
dS_{K}=\frac{8\pi}{4G}\cosh{\left(K  \frac{\pi r_a^2}{G}  
\right)}r_{a}\dot{r_{a}}dt.
\ee
Inserting equations (\ref{Horentropy}),(\ref{Th}),  and 
(\ref{dsk}) into the first 
law of thermodynamics,  and substituting $\dot r_{a}$ using (\ref{apphor}), we 
obtain
\be \label{gfe1}
-4\pi G(\rho_{m}+p_{m})=\cosh{\left[K  
\frac{\pi}{G(H^2+\frac{k}{a^2})}\right]}\left (\dot{H}-\frac{k}{a^2} \right ).
\ee
Finally, inserting the matter conservation equation into (\ref{gfe1}) 
and integrating,   we obtain
\begin{eqnarray} \label{gfe2}
&&
\!\!\!\!\!\!\!\!\!\!\!\!\!\!\!\!
\frac{8\pi G}{3}\rho_{m}= \cosh{\left[K  
\frac{\pi}{G(H^2+\frac{k}{a^2})}\right]}\left (H^{2}+\frac{k}{a^2} \right 
)
\nonumber\\
&&\ \ \ \ \ 
-\frac{K\pi}{G} \text{shi}{\left[K  
\frac{\pi}{G(H^2+\frac{k}{a^2})}\right]}-\frac{\Lambda}{3},
\end{eqnarray}
 where $\Lambda$ is the integration constant and 
$\text{shi}{(x)}$\footnote{The function $\text{shi}{(x)}$ is defined in 
general as $\text{shi}{(x)}=\int^{x}_{0}{\frac{\sinh(x')}{x'}dx'}$.} an entire 
mathematical odd function of $x$ with no branch   discontinuities.

Equations (\ref{gfe1}) and (\ref{gfe2}) are  the modified Friedmann equations, 
obtained by the use of generalized Kaniadakis entropy in the first law of 
thermodynamics, which contain extra terms comparing to the standard 
cosmological equations of General Relativity. As expected, for $K=0$ the 
modified equations (\ref{gfe1}) and (\ref{gfe2}) reduce to the standard ones. 

Moreover, focusing on the flat case, namely $k=0$, we can rewrite the above 
equations as
\begin{eqnarray}
\label{FR1}
&&H^2 = \frac{8\pi G}{3}(\rho_{m}+\rho_{DE})\\
&&\dot{H} = -4\pi G(\rho_{m}+p_{m}+\rho_{DE}+p_{DE}),
\label{FR2}
\end{eqnarray}
 where the dark energy sector is defined as
\begin{eqnarray}
&&
\!\!\!\!\!\!\!\!\!\!\!\!\!\!\!\!\!\!\!\!\!\!\!\!\!\!\!\!\!\!\!\!\!\!
\rho_{DE}=\frac{3}{8\pi G}\left \{\frac{\Lambda}{3}+H^{2}\left [1- 
 \cosh{\left(K  
\frac{\pi}{GH^2}\right)}\right ]\right.\nonumber\\
&& \!\!\!\!\!\!\!  \!\!\!\!\!\!\!\!\! \left. +\frac{K\pi}{G} 
\text{shi}{\left(K \frac{\pi}{GH^2}\right)} \right \},
\label{rhoDE1}
\end{eqnarray}
\begin{eqnarray}
&& \!\!\!\!\!\!\!\!\!\! 
p_{DE}=-\frac{1}{8\pi G}\Big \{\Lambda +(3H^{2}+2\dot{H})\left [1- 
\cosh{\left(K  
\frac{\pi}{GH^2}\right)}\right ] \ \nonumber\\
&&
\ \ \ \ \, 
 +\frac{3K\pi}{G} \text{shi}{\left(K \frac{\pi}{GH^2}\right)} \Big
\}.
\label{pDE1}
\end{eqnarray}
 Hence, with the effective dark energy density
and pressure at hand, we can define the equation-of-state parameter for the 
effective dark energy sector as
\begin{eqnarray}
\label{wDE}
&&
\!\!\!\!\!\!\!\!\!\!\!\!\!\!\!\!\!\!\!\!\!\!
w_{DE}\equiv\frac{p_{DE}}{\rho_{DE}}=-1-
 2\dot{H}\left [1-\cosh{\left(K  
\frac{\pi}{GH^2}\right)}\right ] \nonumber\\
&&
\ \ \ \ \ \ \ \ \
\cdot
\left\{\Lambda +3H^{2}\left [1-\cosh{\left(K  
\frac{\pi}{GH^2}\right)}\right 
] \right.
\nonumber\\
&&
\ \ \ \ \ \ \ \ \ \ \ \
\left.
+\frac{3K\pi}{G} \text{shi}{\left(K  
\frac{\pi}{GH^2}\right)}\right\}^{-1}. 
\end{eqnarray}
  It is clear that in the case where $K=0$, the generalized Friedmann equations 
(\ref{FR1}),(\ref{FR2}) reduce to the standard $\Lambda$CDM cosmology. 
Equations 
 (\ref{FR1}) and (\ref{FR2}) are the modified 
cosmological equations of the scenario at hand, and can determine the evolution 
of the Universe which is being examined in the next section.

\section{Cosmic evolution} \label{cevol}

The constructed modified scenario of the previous section, namely cosmological 
equations   (\ref{FR1}) and (\ref{FR2}), will constitute the pillar in our 
investigation of the cosmological evolution of the Universe. Since we are 
interested in providing analytical solutions too, we focus on the case of dust 
matter, namely we impose $p_{m}=0$. In this case the matter conservation 
equation leads to 
 $\rho_{m} = \frac{\rho_{m0}}{a^3}$, where $\rho_{m0}$ is the value of the 
matter 
energy 
density at the current scale factor which is set to $a_0=1$ (in what follows 
the subscript ``0" 
will denote the 
present value of a quantity).

At this point, it proves convenient to introduce the dimensionless parameters 
\begin{eqnarray} \label{omatter}
&&\Omega_m=\frac{8\pi G}{3H^2} \rho_m\\
&& \label{ode}
\Omega_{DE}=\frac{8\pi G}{3H^2} \rho_{DE},
 \end{eqnarray} 
  for the matter and dark energy density sector respectively. Furthermore, 
equation (\ref{omatter}) gives 
immediately 
$\Omega_m=\Omega_
{m0} H_{0}
^2/a^3 H^2$ and recalling the fact that $\Omega_m + \Omega_{DE}=1$ we can 
obtain an expresssion for the Hubble parameter which reads as 
\be \label{h2}
H=\frac{\sqrt{\Omega_{m0}} H_{0}}{\sqrt{a^3 (1-\Omega_{DE})}}.
\ee
In what follows we will use the redshift $z$ as the independent variable 
($1+z=1/a$ for $a_0=1$). Thus, differentiating (\ref{h2}) we obtain 
\be \label{hddot}
\dot H=-\frac{H^2}{2(1-\Omega_{DE})}[3(1-\Omega_{DE})+(1+z)\Omega'_{DE}],
\ee
 where prime denotes derivative with respect to $z$. This relation
  will be used to eliminate $\dot{H}$ from the above equations.

In order to provide analytical 
solutions, it proves convenient to perform  Taylor expansions of $\cosh(x)$ and 
$\text{shi}(x)$ for small $K$, which is indeed   the case since 
modified Kaniadakis entropy is expected to be close to the standard 
Bekenstein-Hawking one. Hence, 
using that $
 \cosh{(x)}=1+\frac{x^2}{2}+\frac{x^4}{24}+\dots$ and $
\text{shi}{(x)}=x+\frac{x^3}{18}+\frac{x^5}{600}+\dots
$,
expanding the first Friedmann equation and 
using (\ref{h2}), we obtain
\begin{eqnarray} \label{eqomegadecc}
&&\left [\frac{\Lambda}{3H^{2}_{0}\Omega_{m0}(1+z)^{3}}+1 \right 
][1-\Omega_{DE}(z)]
\nonumber\\
&&
+\frac{\pi^{2}K^{2}}{2G^{2}H^{4}_{0}\Omega^{2}_{m0}(1+z)^{6}}
[1-\Omega_{DE}(z)]^{2}\nonumber\\
&&
+\frac{\pi^{4}K^{4}}{18G^{4}H^{8}_{0}\Omega^{4}_{m0}(1+z)^{12
}}[1-\Omega_{DE}(z)]^{4}\approx1.
\end{eqnarray}
 Moreover, applying (\ref{eqomegadecc}) at present time, namely $z=0$, provides 
the modified scenario with a relation between the two free parameters $K$ and 
$\Lambda$, which reads as 
\be \label{fixedcc} 
\Lambda = 3H^{2}_{0}(1-\Omega_{m0})-\frac{\pi^{2}K^{2}}{2G^{2}H^{2}_{0}}\left 
(3+\frac{\pi^{2}K^{2}}{3G^{2}H^{4}_{0}}\right ),
\ee
leaving the scenario with one free parameter, as one can eliminate one of the 
two   parameters in terms of the observationally determined quantities 
$\Omega_{m0}$ and $H_{0}$. Note that for $K\rightarrow 0$, all the above 
obtained equations reduce to the ones of   $\Lambda$CDM cosmology.

Substituting (\ref{fixedcc}) into (\ref{eqomegadecc}) we obtain the solutions 
for $\Omega_{DE}(z)$, which  read as
\begin{eqnarray} 
 \label{omegaDEcc}
&&
\!\!\!\!\!\!\!\!\!\!\!\!\!\!\!\!\!\!
\Omega_{DE}(z)=
1+\frac{\epsilon_{1}}{2}\!\left 
[\frac{3}{\mathcal{A}^2}\mathcal{C}\!-\!\frac{6}{\mathcal{A}}\!-\!
\frac{5}{\mathcal{C}}
\right ]^{1/2} 
\nonumber\\
&&\!\!\!\!\!\!\!\!\!\!\!\!\!\!\!\!\!\!
+\frac{\epsilon_{2}}{2}\!\left 
[\frac{12}{\mathcal{A}}\!-\!\frac{5}{\mathcal{C}}\!+\!\frac{3}{\mathcal{A}^2}
\mathcal{ C }\!-\!\frac{36\mathcal{B}}
{\mathcal{A}^{2}\left 
[\frac{3}{\mathcal{A}^2}\mathcal{C}\!-\!\frac{6}{\mathcal{A}}\!-\!
\frac{5}{\mathcal{C}}
\right ]^{1/2}}\right ]^{1/2}\!,
\end{eqnarray}
 with 
\begin{eqnarray}
\nonumber &&
\!\!\!\!\!\!\!\!\!\!\!\!\!\!\!\!\!\!
\mathcal{A}=\frac{K^2 \pi^2}{G^2 H^{4}_{0}\Omega^{2}_{m0}(1+z)^6}, \\ \nonumber
&& \!\!\!\!\!\!\!\!\!\!\!\!\!\!\!\!\!\!\!\!
\mathcal{B}= 
1+\frac{1-\Omega_{m0}}{\Omega_{m0}(1+z)^3}\\ \nonumber
&& \!\!\!\!\!\!\!\!\!\!\!\!\!\!\!\!\!\!\!\!
-\frac{1}{2}\mathcal{A}\Omega_{m0}(1+z)^{3}(1+\frac{1}{9}\mathcal{A}\Omega^{2}_{m0}(1+z)^{6}), \\ \nonumber
&& \!\!\!\!\!\!\!\!\!\!\!\!\!\!\!\!\!\!\!\!
\mathcal{C}=\!\left 
[9\mathcal{A}^{3}
\!+\!6\mathcal{B}^{2}\mathcal{A}^{2}\!+\!\mathcal{A}^{2} 
\sqrt{\frac{125 }{27}\mathcal{A
}^{2}\!+\!36
\left(\frac{3}{2}\mathcal{A}\!+\!\mathcal{B}^{2}
\right)^ { 2}}\right ]^{\frac{1}{3}}\!.
\end{eqnarray}
and where $\epsilon_{1}, \epsilon_{2}=\pm1$.
Finally, differentiating (\ref{omegaDEcc}) and inserting into  
(\ref{h2}),(\ref{hddot}) and then into (\ref{wDE}) we can obtain the analytical 
expression for the   dark energy equation-of-state parameter 
$w_{DE}(z)$. Lastly, the other physically interesting quantity, namely 
 the deceleration parameter $q\equiv -1-\frac{\dot H}{H^{2}}$ can be similarly 
calculated using (\ref{h2}),(\ref{hddot}) and the solution (\ref{omegaDEcc}).

In conclusion, we were able to extract analytical solutions for the observable 
quantities of the dark energy sector, namely for $\Omega_{DE}$, $w_{DE}$ and 
$q$, of the constructed cosmological scenarios through Kaniadakis entropy. In 
the following subsections we investigate in more detail their cosmological 
implications.

\subsection{$\Lambda = 0$ case}

We start our analysis from the case where  an explicit 
cosmological constant is absent. We mention that in this case the scenario at 
hand does not have $\Lambda$CDM cosmology as a limit, i.e. it corresponds to a 
  radical modification of standard cosmology with extra terms depending on the 
Kaniadakis exponent $K$. 

In the absence of $\Lambda$, the dark-energy sector relations 
(\ref{rhoDE1}), (\ref{pDE1}) and (\ref{wDE}) respectively become
\begin{eqnarray}
&&
\!\!\!\!\!\!\!\!\!\!\!\!\!\!\!\!\!\!\!\!\!\!\!\!\!\!\!\!\!\!\!\!\!\!\!\!\!\!\!\!
\rho_{DE}=\frac{3}{8\pi G}\left \{ H^{2}\left [1- 
 \cosh{\left(K  
\frac{\pi}{GH^2}\right)}\right ]\right.\nonumber\\
&& \!\!\! \!\left. +\frac{K\pi}{G} 
\text{shi}{\left(K \frac{\pi}{GH^2}\right)} \right \},
\label{rhoDE1ccnot}
\end{eqnarray}
\begin{eqnarray}
&& \!\!\!\!\!\!\!\!\!\! \!\!\!\!\!
p_{DE}=-\frac{1}{8\pi G}\Big \{ (3H^{2}+2\dot{H})\left [1- 
\cosh{\left(K  
\frac{\pi}{GH^2}\right)}\right ] \ \nonumber\\
&&
\ \ \ \ \, \ \ \ \ \,  \ \ \ \, 
 +\frac{3K\pi}{G} \text{shi}{\left(K \frac{\pi}{GH^2}\right)} \Big
\}.
\label{pDE1ccnot}
\end{eqnarray} 
and
\begin{eqnarray}
\label{wDEccnot}
&&
\!\!\!\!\!\!\!\!\!\!\!\!\!\!\!\!\!\!\!\!\!\!
w_{DE}\equiv\frac{p_{DE}}{\rho_{DE}}=-1-
 2\dot{H}\left [1-\cosh{\left(K  
\frac{\pi}{GH^2}\right)}\right ] \nonumber\\
&&
\ \ \ \ \ \ \ \ \ \,
\cdot
\left\{ 3H^{2}\left [1-\cosh{\left(K  
\frac{\pi}{GH^2}\right)}\right 
] \right.
\nonumber\\
&&
\ \ \ \ \ \ \ \ \ \ \ \
\left.
+\frac{3K\pi}{G} \text{shi}{\left(K  
\frac{\pi}{GH^2}\right)}\right\}^{-1}. 
\end{eqnarray}
However,   when we apply  (\ref{eqomegadecc}) at 
present time, instead of  (\ref{fixedcc}) in the case of $\Lambda=0$  we 
acquire 
\begin{eqnarray}\label{relK0} 
3H^{2}_{0}(1-\Omega_{m0})=\frac{\pi^{2}K^{2}}{2G^{2}H^{2}_{0}}\left 
(3+\frac{\pi^{2}K^{2}}{3G^{2}H^{4}_{0}}\right ).
\end{eqnarray}
Hence, in the absence of 
$\Lambda$ the parameter $K$ is not completely free but it should vary in a 
range consistent with the observational range of $\Omega_{m0}$, and of course 
the case $K=0$ is now excluded since it corresponds to dark-energy absence (as 
we mentioned above the scenario at hand does not have $\Lambda$CDM cosmology as 
a limit, and in the case  $K=0$ it gives just CDM scenario).

The solution for the dark energy density parameter 
from (\ref{eqomegadecc}) is still given by (\ref{omegaDEcc})
 but with
\begin{eqnarray}
\nonumber &&
\!\!\!\!\!\!\!\!\!\!\! 
\mathcal{A}=\frac{K^2 \pi^2}{G^2 H^{2}_{0}\Omega^{2}_{m0}(1+z)^6}, \\ \nonumber
&& \!\!\!\!\!\!\!\!\!\!\! 
\mathcal{B}= 1, \\ \nonumber
&& \!\!\!\!\!\!\!\!\!\!\! 
\mathcal{C}=\!\left 
[9\mathcal{A}^{3}
\!+\!6
\mathcal{A}^{2}\!+\!\mathcal{A}^{2} 
\sqrt{\frac{125 }{27}\mathcal{A
}^{2}\!+\!36
\left(\frac{3}{2}\mathcal{A}\!+\! 1
\right)^ { 2}}\right ]^{\frac{1}{3}}\!.
\end{eqnarray} 
 \begin{figure}[!h]
\centering
\includegraphics[width=7.25cm]{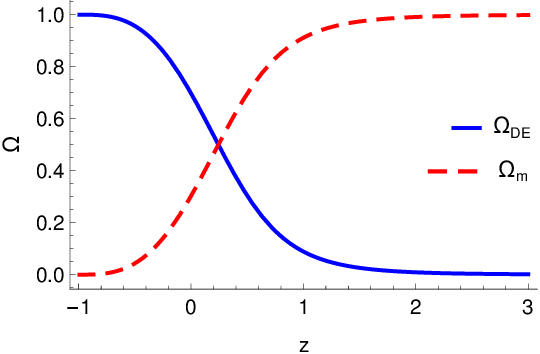}

\includegraphics[width=7.25cm]{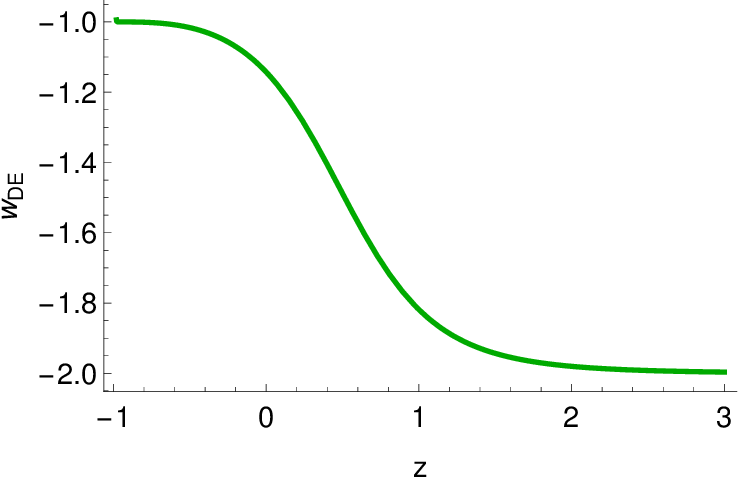} 
\includegraphics[width=7.25cm]{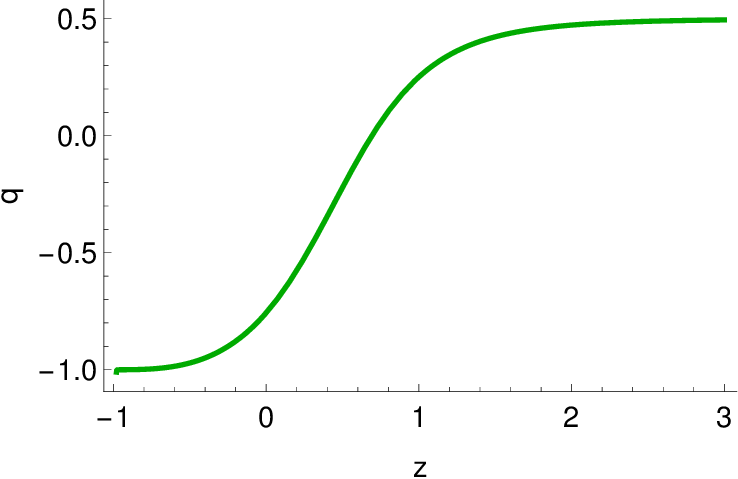}
\caption{
\it{{\bf{Upper   graph}}: 
The evolution of the  effective dark 
energy density parameter $\Omega_{DE}$} (blue-solid) and the
matter density parameter $\Omega_{m}$ (red-dashed) respectively, as a 
function of the redshift $z$, for the
modified scenario through Kaniadakis entropy, in the case of $\Lambda=0$.
    {\bf{ Middle graph}}: 
The evolution of the   effective dark energy equation-of-state 
parameter $w_{DE}$. {\bf{ Lower   graph}}: Evolution of the   
deceleration parameter $q$. In all graphs the Kaniadakis entropic parameter is 
$K=0.35$ in units of $k_{_B}$,  and   according to 
(\ref{relK0}) $ \Omega_{m0} \approx0.3$. }
\label{OmegasL0}
\end{figure}

In the upper  graph of Fig. \ref{OmegasL0}  we present the 
evolution of the physically accepted energy densities $\Omega_{DE}$ and 
$\Omega_{m}$  in the case where $K=0.35$ (in units of $k_B$), which 
according to (\ref{relK0}) corresponds to  $\Omega_{m} (z = 0) = 
\Omega_{m0} \approx 0.30$.   
As we 
can see,  we 
acquire the usual thermal history of the Universe, with the sequence of matter 
and dark-energy epochs,   while in the asymptotic future  
the Universe results in a dark-energy dominated, de Sitter phase.
However,   the dark-energy equation-of-state parameter $w_{DE}$, although being 
close to $-1$ at present, and in the future, at large redshifts it goes to 
$-2$. This behavior is inside the observational bounds \cite{Planck:2018vyg}, 
nevertheless it is less attractive. 
Finally, from the deceleration parameter  we can see that the 
transition from deceleration to acceleration takes place at a redshift 
$z_{tr}\approx0.6$, in agreement with   observations.

\begin{figure}[!h]
\centering
\includegraphics[width=8.1cm]{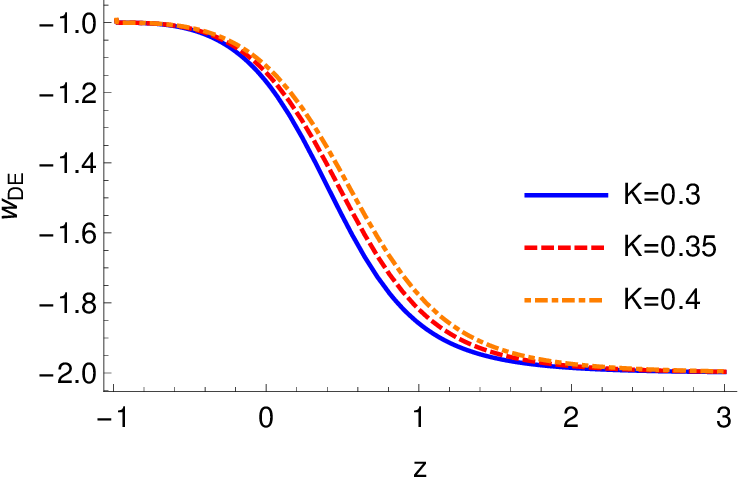}                                   
 
\caption{
\it{The  evolution of the effective dark energy equation-of-state 
parameter $w_{DE}$, for different values of the Kaniadakis entropic parameter 
$K$ in unitsof $k_{_B}$, in the case of $\Lambda=0$. In all cases we have 
obtained density parameters 
evolution similar to the graphs of Fig.\ref{OmegasL0}, and $\Omega_{m0}$ lies 
inside the $2\sigma$ region according to 
Planck Collaboration,  namely $ 
\Omega_{m0}\approx0.31\pm0.014$ \cite{Planck:2018vyg}.} 
}
\label{fig:multiwde22}
\end{figure}

Let us now study in more detail the effect of the entropic parameter  $K$ on 
the 
cosmic evolution and in particular focusing on the dark energy 
equation-of-state parameter. In Fig. \ref{fig:multiwde22} we depict $w_{DE}$ 
for 
different values of  $K$. The behavior is similar to the one described above, 
namely $w_{DE}$ starts from -2, and it becomes around -1 at present and future, 
while lying always in the phantom regime. Note that the transition redshift has 
a slight dependence on $K$. Finally, note that in order to have $ 
\Omega_{m0}\approx0.31\pm0.014$, which is the $2\sigma$ region according to 
Planck Collaboration \cite{Planck:2018vyg}, $K$ is varied in the range 
$0.3\lesssim K \lesssim  0.45$.
Lastly, at asymptotically large times    the Universe 
  results always in a dark-energy dominated, de-Sitter phase.

\subsection{$\Lambda \neq 0$ case}

In the previous subsection we examined the case where an explicit cosmological 
constant is absent, and as we saw the obtained  results although in agreement 
with observation were not completely attractive since the limit $K\rightarrow0$ 
could not be obtained and moreover the early-time behavior of $w_{DE}$ was 
around -2. Hence, in this subsection we consider the case where an explicit 
cosmological constant is present, namely we consider     $\Lambda \neq 
0$.  In this case, for $K=0$ the scenario does give back  
$\Lambda$CDM cosmology, nevertheless for $K \neq 0$ the extra terms due to 
Kaniadakis entropy trigger    deviations from $\Lambda$CDM scenario, which is 
exactly the focus of interest of the present work.

\begin{figure}[h!]
\centering
\includegraphics[width=7.25cm]{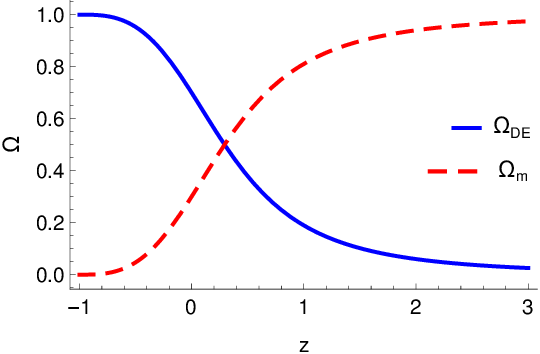}                                      
\includegraphics[width=7.25cm]{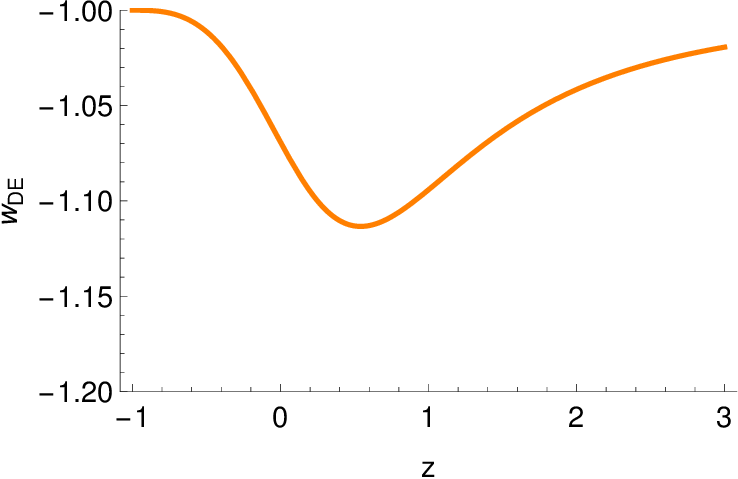} 
\includegraphics[width=7.25cm]{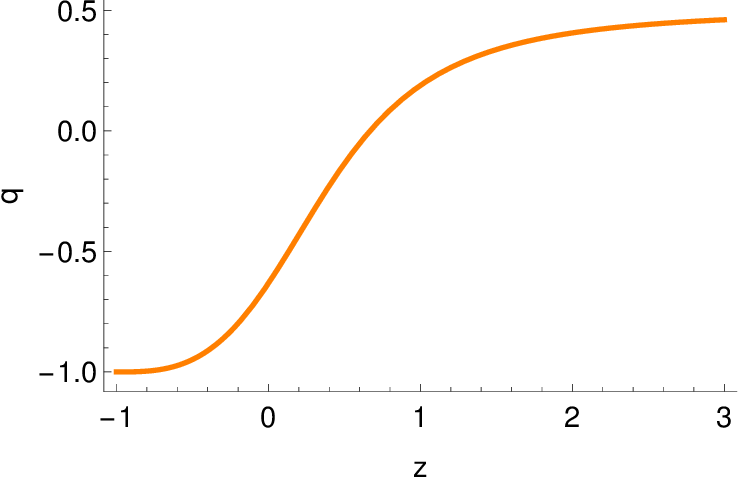}
\caption{
\it{{\bf{Upper   graph}}: 
The evolution of the  effective dark 
energy density parameter $\Omega_{DE}$} (blue-solid) and the
matter density parameter $\Omega_{m}$ (red-dashed) respectively, as a 
function of the redshift $z$, for the
modified scenario through Kaniadakis entropy.
    {\bf{ Middle graph}}: 
The evolution of the   effective dark energy equation-of-state 
parameter $w_{DE}$. {\bf{ Lower   graph}}: Evolution of the   
deceleration parameter $q$. In all graphs the Kaniadakis entropic parameter is 
$K=0.2$ in units of $k_{_B}$,  while $\Lambda$ is given by 
(\ref{fixedcc}), and we have fixed   $\Omega_{m} (z = 0) = \Omega_{m0} \approx 
0.3$.}
\label{OmegasL}
\end{figure}

In the upper  graph of Fig. \ref{OmegasL}  we depict the 
evolution of the energy densities $\Omega_{DE}$ and $\Omega_{m}$, as given by 
the analytical solution
(\ref{omegaDEcc})\footnote{From the four solutions of $\Omega_{DE}$ we keep 
only 
the solution with sign ($+,+$), while we discard the other three,  since they 
lead either to early-time dark energy, or to negative $\Omega_{DE}$, 
or to the wrong sequence of matter and dark energy epochs.} 
and by
$\Omega_{m}(z)=1-\Omega_{DE}(z)$ respectively, in the case where $K=0.2$. 
Note that we impose $\Omega_{m} (z = 0) = 
\Omega_{m0} \approx 0.3$ in agreement with the 
  Planck results  \cite{Planck:2018vyg}. 
As we 
observe, from the resulting evolution of $\Omega_{DE}$ and $\Omega_{m}$ we 
obtain the usual thermal history of the Universe in agreement with 
observations, while in the asymptotic future ($z\rightarrow-1$) the Universe 
results in a dark-energy dominated, de Sitter phase.

Additionally, in the middle graph of Fig. \ref{OmegasL} we present the 
evolution of the dark-energy equation-of-state parameter $w_{DE}$. As can 
be seen it slightly lies  in the phantom regime throughout the Universe 
evolution, nevertheless still inside the observational bounds 
\cite{Planck:2018vyg}, while in the asymptotic future it goes to de Sitter 
phase as mentioned above.  Lastly, in the lower graph we depict the 
corresponding deceleration parameter $q(z)$. From this plot we can see the 
transition from deceleration to acceleration at a redshift $z_{tr}\approx0.6$, 
in agreement with the observed behavior.

\begin{figure}[!h]
\centering
\includegraphics[width=8.1cm]{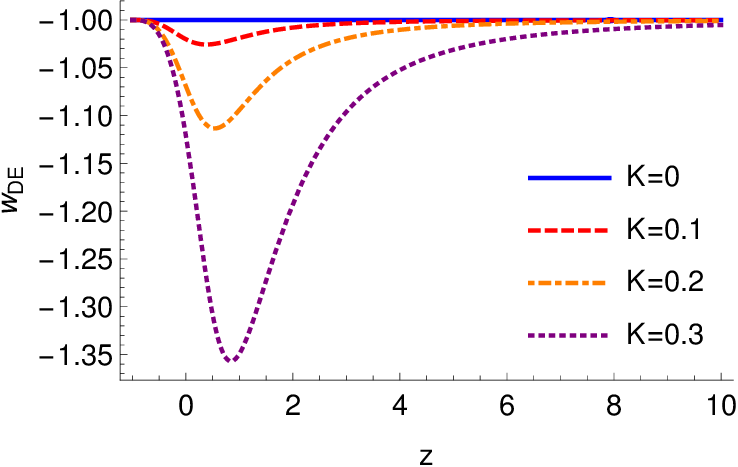}                                
  \caption{\it{The  evolution of the effective dark energy equation-of-state 
parameter $w_{DE}$, for different values of the Kaniadakis entropic parameter 
$K$. We have imposed $\Omega_{m0} \approx 0.3$ at present time in units of 
$k_{_B}$, and in all cases we have obtained density parameters 
evolution similar to the graphs of Fig.\ref{OmegasL}.
}}
\label{fig:multiwdeL}
\end{figure}

We proceed by examining  the effect of the entropic parameter  
$K$ 
on the    dark energy 
equation-of-state parameter. In Fig. \ref{fig:multiwdeL} we present $w_{DE}$ 
for 
different values of Kaniadakis parameter $K$. As we stated above, for 
$K\rightarrow 0$ we re-obtain the $\Lambda$CDM scenario, i.e. 
$w_{DE}=-1=const.$. 
As the Kaniadakis parameter increases, the dark energy shows a dynamical 
behaviour, with $w_{DE}$ at larger redshifts   lying slightly in the phantom
regime, 
but at small redshifts and current time  it deviates more significantly from 
$\Lambda$CDM cosmology. 
Finally, at asymptotically large times,  it will 
always stabilize at the cosmological constant value $-1$, and the Universe 
always results in the de-Sitter solution, independently of the Kaniadakis 
parameter $K$. Note that  $w_{DE}$ is always in the phantom regime, which is an 
advantage of the scenario, since it is known that the phantom regime cannot be 
easily obtained.

We close this subsection with the calculation of the Universe age
according to the scenario at hand.
 Starting from the expression
$
t(z)=\int^{\infty}_{z}\frac{dz^{'}}{(1+z^{'})H(z^{'})}
$
and inserting a typical $H(z)$ evolution obtained above we find 
$t(z)=\frac{0.966279}{H_{0}}$, and thus with the Planck value $H_0 = (67.27\pm 
0.60)$ km/s/Mpc   we finally obtain
\be 
t_{age}=13.936^{+0.017}_{-0.017}\ Gyrs.
\ee
This value coincides within 1$\sigma$ with the value corresponding to 
$\Lambda$CDM scenario, namely  $13.787^{+0.020}_{-0.020}$ Gyrs 
\cite{Planck:2018vyg}.

\subsection{Relation with new Tsallis entropy}

For completeness, in this subsection we examine the relation of Kaniadakis 
entropy with New Tsallis entropy. The latter can be written as
\be \label{ntentropy}
S_{T}=\frac{2\exp\left (\frac{\delta S_{BH}}{2}\right )}{\delta}\sinh{\left 
(\frac{\delta S_{BH}}{2}\right )},
\ee
and as it can be seen at small $\delta$ it is quite similar with Kaniadakis 
entropy (\ref{kentropy}).
Repeating the steps of subsection \ref{mfekentropy}, but using the above 
entropy instead of Kaniadakis entropy we obtain the following 
modified Friedmann equations:
\begin{eqnarray} \label{ntgfe1}
&&\!\!\!\!\!\!\!
-4\pi G(\rho_{m}+p_{m})=e^{\frac{\delta \pi}{G(H^2+\frac{k}{a^2})}}\left 
(\dot{H}-\frac{k}{a^2} \right ) \nonumber \\ 
&&\!\!\!\!\!\!\!
\cdot \left \{\sinh{\left [\frac{\delta \pi}{G(H^2+\frac{k}{a^2})} \right ]}+\cosh{\left [\frac{\delta \pi}{G(H^2+\frac{k}{a^2})}\right \}}\right ].
\end{eqnarray}
\begin{eqnarray} \label{ntgfe2}
\frac{8\pi G}{3}\rho_{m}&=& e^{\frac{\delta \pi}{G(H^2+\frac{k}{a^2})}}\left 
(H^{2}+\frac{k}{a^2} \right )\nonumber\\ 
&& 
-\frac{\delta \pi}{G} \text{Ei}{\left [\frac{\delta \pi}{G(H^2+\frac{k}{a^2})}\right ]}-\frac{\Lambda}{3},
\end{eqnarray}
where
$\text{Ei}{(x)}$ is defined as 
$\text{Ei}{(x)}=-\int^{\infty}_{-x}{\frac{e^{-x'}}{x'}dx'}$.
Additionally, the effective dark energy density and pressure become
\begin{eqnarray}
\rho_{DE}=\frac{3}{8\pi G}\!\left [\frac{\Lambda}{3}+H^{2}\!\left (1\!-\! 
 e^{\frac{\delta \pi}{GH^2}}\right )\! +\frac{\delta \pi}{G} 
\text{Ei}{\left(\frac{\delta \pi}{GH^2}\right)} \!\right ],
\label{ntrhoDE1}
\end{eqnarray}
\begin{eqnarray}
&&\!\!\!\!\!\!\!
\!\!\!\!\!\!\!\!\!\!\!\!\!\!\!\!\!\!\!\!\!\!\! 
p_{DE}=-\frac{1}{8\pi G} \bigg [\Lambda +(3H^{2}+2\dot{H})\left 
(1- 
 e^{\frac{\delta \pi}{GH^2}}\right )  \\ \nonumber
&& \ \ \ \ \,
 +\frac{3\delta \pi}{G} \text{Ei}{\left(\frac{\delta \pi}{GH^2}\right)}  \bigg
].
\label{ntpDE1}
\end{eqnarray}
Expanding the $\text{Ei}(x)$ function for small $\delta$ as $
\text{Ei}{(x)}=\gamma +\log(x)+x+\frac{x^2}{4}+\dots$, where $\gamma$ is 
Euler's constant, we finally obtain
\be 
\Omega_{DE}=1-\frac{-\mathcal{D}-\mathcal{E}\mathcal{F}+\sqrt{4\mathcal{E}\left 
(1+\mathcal{E}\right )+\left (\mathcal{D}+\mathcal{E}\mathcal{F}\right 
)^2}}{2\mathcal{E}\left (1+\mathcal{E}\right )},
\ee
with 
\begin{eqnarray}
\nonumber &&
\!\!\!\!\!\!\!\!\!\!\! 
\mathcal{D}=\frac{\Lambda}{3H^{2}_{0}\Omega_{m0}(1+z)^3}+1, \\ \nonumber
&& \!\!\!\!\!\!\!\!\!\!\! 
\mathcal{E}= \frac{\delta \pi}{GH^{2}_{0}\Omega_{m0}(1+z)^3}, \\ \nonumber
&& \!\!\!\!\!\!\!\!\!\!\! 
\mathcal{F}=\gamma +\log(\mathcal{E})-2.
\end{eqnarray} 
Lastly, applying the first Friedmann equation at present we extract the 
relation between the two free parameters $\delta$ and 
$\Lambda$, namely
\begin{eqnarray} \label{ntfixedcc}
&&\!\!\!\!\!\!\!\!\!\!\!\!
\Lambda = 3H^{2}_{0}(1-\Omega_{m0})-\frac{\delta^{2} 
\pi^{2}}{G^{2}H^{2}_{0}} \nonumber\\
&&
-\frac{3\delta \pi}{G}\left [\gamma +\log\left ({\frac{\delta 
\pi}{G\Omega_{m0}H^{2}_{0}}}\right ) \Omega_{m0}-2\right ],
\end{eqnarray}
leaving the scenario with one free parameter.  Note that for $\delta 
\rightarrow 0$, all the above obtained equations reduce to the ones of 
$\Lambda$CDM cosmology.

\begin{figure}[!h]
\centering
\includegraphics[width=8.1cm]{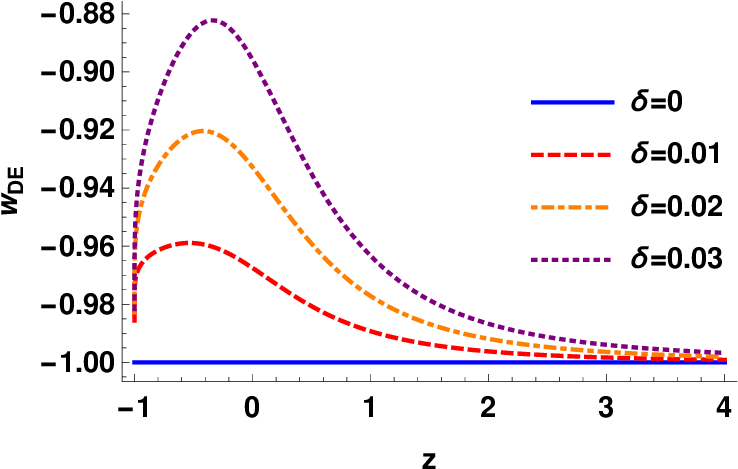}                                    
\caption{\it{The  evolution of the effective dark energy equation-of-state 
parameter $w_{DE}$, for different values of the new Tsallis entropy parameter 
$\delta$ in (\ref{ntentropy}), in the case of $\Lambda \neq 0$.}}
\label{fig:ntmultiwde22}
\end{figure}

Elaborating the above equations numerically, we find that the model can indeed 
describe the thermal history of the Universe, with dark-energy density 
parameter, deceleration parameter, and dark-energy equation-of-state parameter 
evolution similarly to Fig. \ref{OmegasL}. Additionally, in Fig. 
\ref{fig:ntmultiwde22}  we display $w_{DE}$ 
for 
different values of the Tsallis parameter $\delta$. As we observe, although for 
$\delta=0$ we recover  $\Lambda$CDM scenario, for $\delta$ deviating from zero 
we obtain a dark-energy sector lying in the quintessence regime, with the 
deviations from $\Lambda$CDM cosmology being larger at small redshifts. 
Moreover, at asymptotically large times the dark-energy  equation-of-state 
parameter 
  stabilizes at the cosmological constant value $-1$, and the Universe 
always results in the de-Sitter solution  independently of the Tsallis 
parameter. Lastly,  we mention that, similarly to Kaniadakis case, an 
explicit cosmological constant is required in order to have efficient 
phenomenology.

\section{Conclusions} \label{concl}

In this work  we have constructed new cosmological scenarios by considering the 
widely-known conjecture that thermodynamics is related to gravity. 
In particular, it is known that  one can start from  the first law of 
thermodynamics, applied in the Universe horizon, and result in 
the Friedmann equations. In this procedure one uses the entropy 
relation, namely the Bekenstein-Hawking one  in the case of General 
Relativity  or the modified entropy expression in the case of modified gravity.
Nevertheless, following the above procedure in the reverse way, and applying 
the generalized Kaniadakis hyperbolic entropy, we extracted  modified 
Friedmann 
equations, which contain extra terms   that appear for the first time. 
These new terms   are quantified 
by the  single, new, Kaniadakis entropy parameter $K$ and effectively give rise 
to a  dark energy sector. 
In the 
case  $K=0$, where Kaniadakis entropy becomes the standard Bekenstein-Hawking 
one,  the above effective dark energy becomes a   constant 
and 
$\Lambda$CDM concordance model is re-obtained.  However, in the case where 
deviations of Kaniadakis from Bekenstein-Hawking 
one are switched on, we acquire   very interesting cosmological behavior.

 In order to study this behavior in a more thorough way, we assumed the matter 
sector to be dust, which allowed us to find    analytical 
solutions for the dark energy density parameter, as well as for the dark-energy
equation of state and for 
the deceleration parameter.
As we saw,    the Universe realizes  the sequence of matter and dark-energy 
epochs, while it transits from deceleration to  
  acceleration at $z_{tr}\approx0.6$ in agreement with the observed behavior. 
    Furthermore, when  we consider
an explicit cosmological constant, according to the value of $K$ 
the equation-of-state parameter of 
dark energy deviates from the cosmological constant value at small redshifts, 
while lying always in the phantom regime. Additionally, at asymptotic late times
 it   stabilizes in the cosmological constant 
value $-1$, i.e. the Universe always results in a dark-energy dominated, de 
Sitter phase.

For completeness, we investigated the sub-case  where there is not an explicit 
cosmological constant. In this case the scenario at hand does not have  
$\Lambda$CDM cosmology as a limit, and the evolution is determined solely by 
the extra terms. We extracted analytical solutions  for the dark energy density 
and we showed that even without $\Lambda$
  the new terms can  trigger the    sequence of matter and dark energy eras. 
Furthermore,  the dark energy equation-of-state parameter
  starts from -2 at large redshifts, and it becomes around -1 at present and 
future times, while lying always in the phantom regime, while the transition 
redshift has a slight dependence on $K$. Note that this $w_{DE}$ behavior is 
still inside the observational bounds of Planck Collaboration, 
 since the deviation from $-1$ takes place at quite early 
times, where the observational errors are huge \cite{Planck:2018vyg}.

In conclusion, the modified cosmology obtained from the gravity-thermodynamics 
conjecture through Kaniadakis entropy leads to very interesting Universe 
evolution. It would be both interesting and necessary to perform a full 
observational analysis using data from Supernova type Ia (SNIa),   
Baryon Acoustic Oscillation (BAO), Cosmic Microwave Background (CMB), 
and Hubble parameter measurements, in order to extract constraints on the 
model parameter $K$. Such an investigation will be performed in a forthcoming 
publication.

\end{document}